\documentclass[a4paper]{jpconf}
\usepackage{graphicx}
\begin{document}
\title{Differences between stellar and laboratory reaction cross sections}

\author{Thomas Rauscher}

\address{Department of Physics, University of Basel, 4056 Basel, Switzerland}

\ead{Thomas.Rauscher@unibas.ch}

\begin{abstract}
Nuclear reactions proceed differently in stellar plasmas than in the laboratory due to the thermal effects in the plasma. On one hand, a target nucleus is bombarded by projectiles distributed in energy with a distribution defined by the plasma temperature. The mostly relevant energies are low by nuclear physics standards and thus require an improved description of low-energy properties, such as optical potentials, required for the calculation of reaction cross sections. Recent studies of low-energy cross sections suggest the necessity of a modification of the proton optical potential. On the other hand, target nuclei are in thermal equilibrium with the plasma and this modifies their reaction cross sections. It is generally expected that this modification is larger for endothermic reactions. We show that there is a large number of exceptions to this rule.
\end{abstract}

\section{Introduction}
Stellar plasmas are characterized by their composition and temperature. The energy distribution of nuclei according to the temperature is given by a Maxwell-Boltzmann (MB) function. This function peaks at comparatively low energy and thus the astrophysically relevant interaction energies of nuclei are low by nuclear physics standards. This is a challenge for experimental and theoretical nuclear physics, especially when dealing with charged-particle induced reactions. The thermalization of nuclei is very fast, even compared to reaction timescales. This populates excited states and reactions can proceed from the ground state and the excited states, contrary to reactions in the laboratory which only include targets in the ground state. The resulting change in the effective cross section is called stellar enhancement and usually given by the ratio of the stellar cross section to the ground state one, $f=\sigma^*/\sigma^\mathrm{g.s.}$. The stellar enhancement factor $f$ can only be studied theoretically.

\section{Coulomb suppression of the stellar enhancement factor}
Stellar reaction rates are obtained by folding \textit{stellar} cross sections with the MB distribution of the projectiles. It can be shown \cite{fowl74,holmes76} that \textit{stellar} rates obey reciprocity whereas rates based on laboratory cross sections do not. Since excited states in both the target and the final nucleus contribute, it is easy to see that more transitions beyond the g.s. transitions are involved in the final nucleus of an exothermic reaction compared to its target nucleus \cite{holmes76,kiss08,rau09} and thus $f_\mathrm{endo}>f_\mathrm{exo}$. When studying an astrophysically relevant reaction in the laboratory, $f$ should be as close as possible to Unity. Therefore, it seems advantageous to study exothermic reactions.
\begin{figure}
\includegraphics[width=11cm,angle=-90,clip]{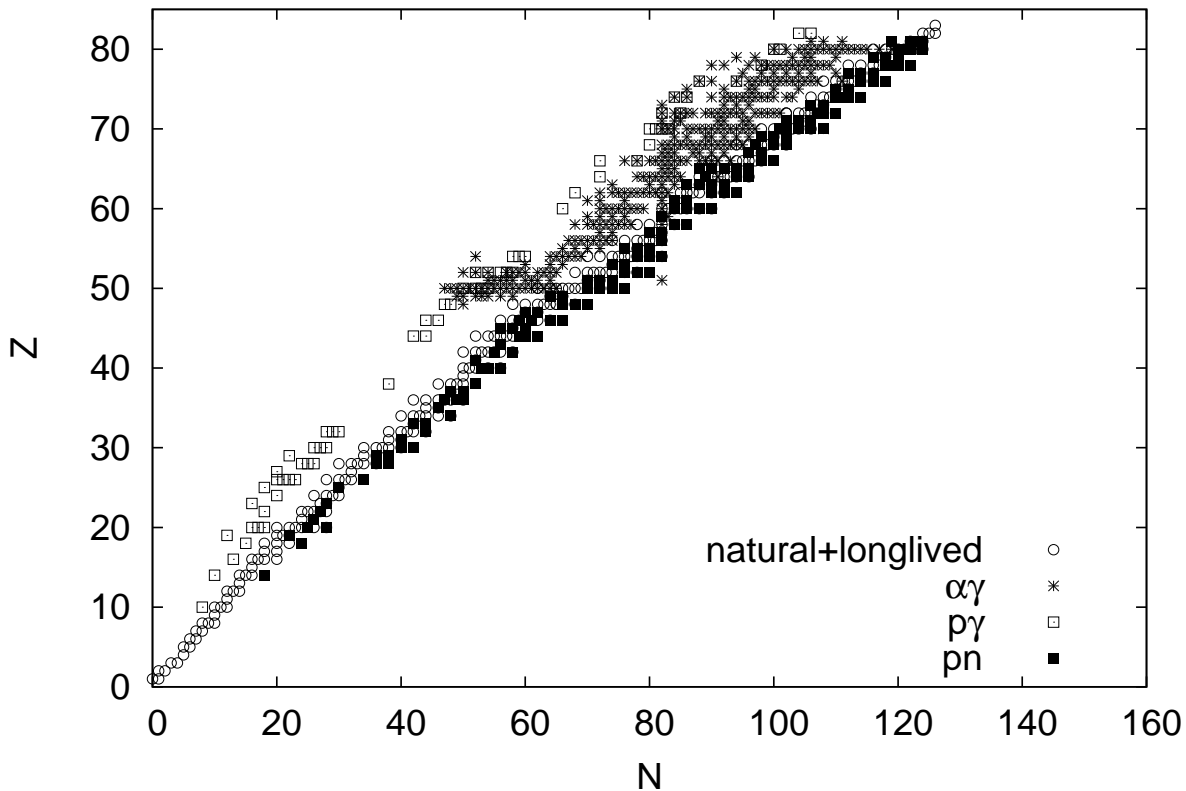}
\caption{\label{fig:supp}Targets for selected reactions with $f_\mathrm{endo}<f_\mathrm{exo}$. Natural and longlived nuclides are shown to guide the eye.}
\end{figure}

Closer inspections reveals \cite{kiss08,rau09} that the above rule is not appropriate in all cases. If transitions from excited state are systematically suppressed, there can be cases for which $f_\mathrm{endo}<f_\mathrm{exo}$ although the endothermic reaction in principle may allow more transitions from excited states. Charged particle compound reactions provide such a suppression mechanism. With increasing excitation energy, the relative energy of the transitions between excited states and the compound state is decreasing. With a sufficiently high Coulomb barrier, the transitions from excited states are efficiently suppressed. We found \cite{kiss08,rau09} that the stellar enhancement factor of the endothermic direction $f_\mathrm{endo}$ is lower than the one for the exothermic direction $f_\mathrm{exo}$ for more than 1200 reactions. This includes some reactions at or close to stability relevant in the p-process and reactions on highly unstable nuclei in the r-, rp-, and $\nu$p-process paths. Examples are shown in Fig.\ \ref{fig:supp}, detailed lists of reactions are given in \cite{rau09}.
\begin{figure}
\begin{minipage}{14pc}
\includegraphics[bb=11bp 108bp 420bp 734bp,clip,angle=-90,width=19pc]{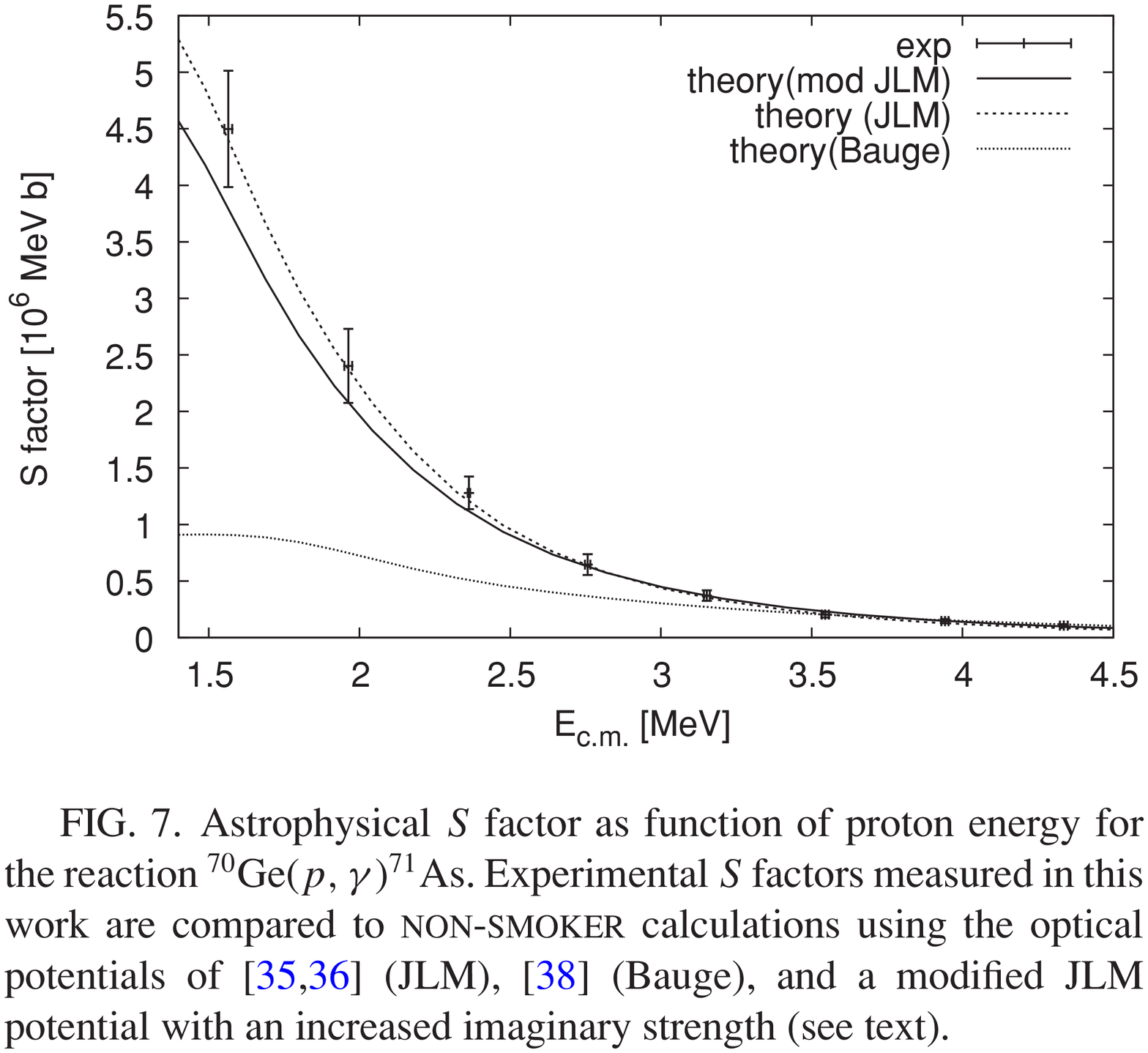}
\end{minipage}\hspace{5.5pc}%
\begin{minipage}{14pc}
\includegraphics[bb=11bp 150bp 450bp 732bp,clip,angle=-90,width=17pc]{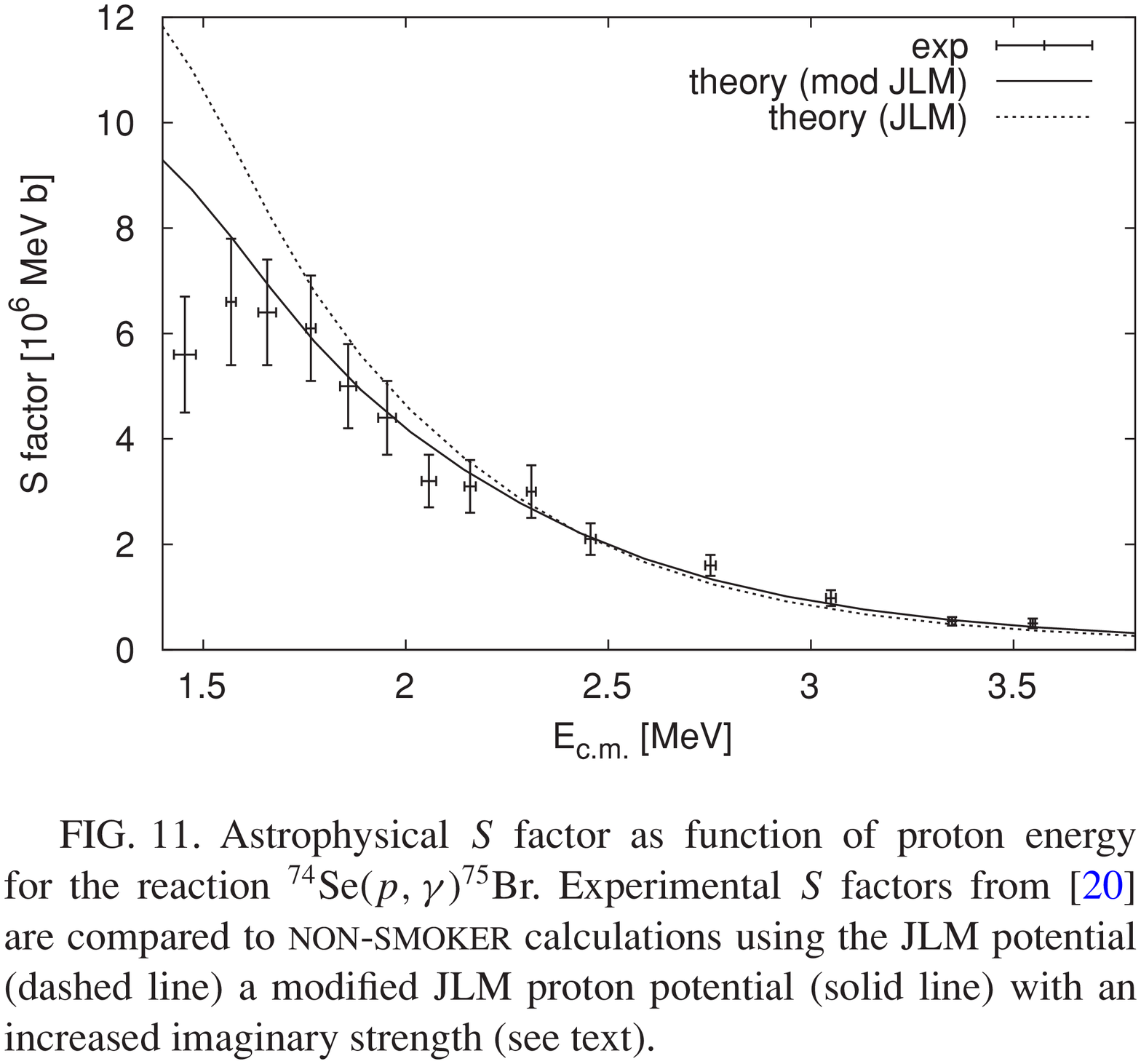}
\end{minipage} 
\caption{\label{fig:pg}Astrophysical S-factors for (p,$\gamma$) reactions on $^{70}$Ge (left) and $^{74}$Se (right; all data are from \cite{prc76kiss}) compared with theoretical values obtained with
different optical potentials: "standard" potential (JLM, \cite{jlm,jlmlow}), our new potential (mod JLM), and the potential by \cite{bauge} (Bauge).}
\end{figure}

\section{The low-energy optical potential for protons}
Measuring low-energy cross sections for charged-particle reactions is problematic due to the Coulomb barrier causing the astrophysically relevant cross sections to be tiny. Even more problematic is the standard way to obtain information on optical potentials through elastic scattering experiments. The scattering cross section at low-energy becomes indistinguishable from Rutherford scattering.  Accordingly, the database for reactions involving protons and $\alpha$s at astrophysically relevant energies is incomplete.

Proton capture rates at $2-4$ GK are important in the $\gamma$-process which is thought to produce the majority of p-nuclei. An important difference to cross sections at higher energies is in the fact that the cross sections of the p-process are mostly sensitive to the particle widths instead of the $\gamma$-widths. This is due to the Coulomb barrier, reducing the charged particle width at low energy to values below those of the $\gamma$-width. A series of (p,$\gamma$) and (p,n) reactions was measured at $\gamma$-process energies recently (see, e.g., \cite{prc76kiss,kiss08} and references therein). The latter reactions are especially useful for testing the proton potential because the neutron width will (almost always) be larger than the proton width at all energies.

Several optical potentials were tested against the data and it was
found that a modification of a widely used microscopic potential reproduces
the data best. This is the potential by
\cite{jlm} with low-energy modifications by \cite{jlmlow} (JLM) which were especially provided for applications in nuclear astrophysics. This microscopic potential is derived by applying the Brueckner-Hartree-Fock approximation with Reid's hard core nucleon-nucleon interaction and adopting a local density approximation.

Despite of overall good agreement, systematic deviations at low energy are still found (see, e.g., Fig.\ \ref{fig:pg}, more examples are shown in \cite{prc76kiss}).
By variation of the optical potential we found that an increase by 70\% in the strength of the imaginary part considerably improved the reproduction of the data in Hauser-Feshbach calculations (denoted by ``mod JLM'' in the figures). This increased absorption is allowed within the previous parameterization because the isoscalar and especially the isovector component of the imaginary part is not well constrained at low energies, as also noted by \cite{jlm,jlmlow,bauge98,bauge,isovec}.

Also shown are results obtained with another recent, Lane-consistent new parameterization of the JLM potential (Bauge \cite{bauge}). It yields worse agreement at low energy but it does neither include the additional modifications of \cite{jlmlow}, nor can it constrain the imaginary isovector part well at low energy \cite{bauge98,bauge,isovec} because
it was fitted to data at higher energy.
\begin{figure}
\begin{minipage}{14pc}
\includegraphics[width=14pc,angle=-90,clip]{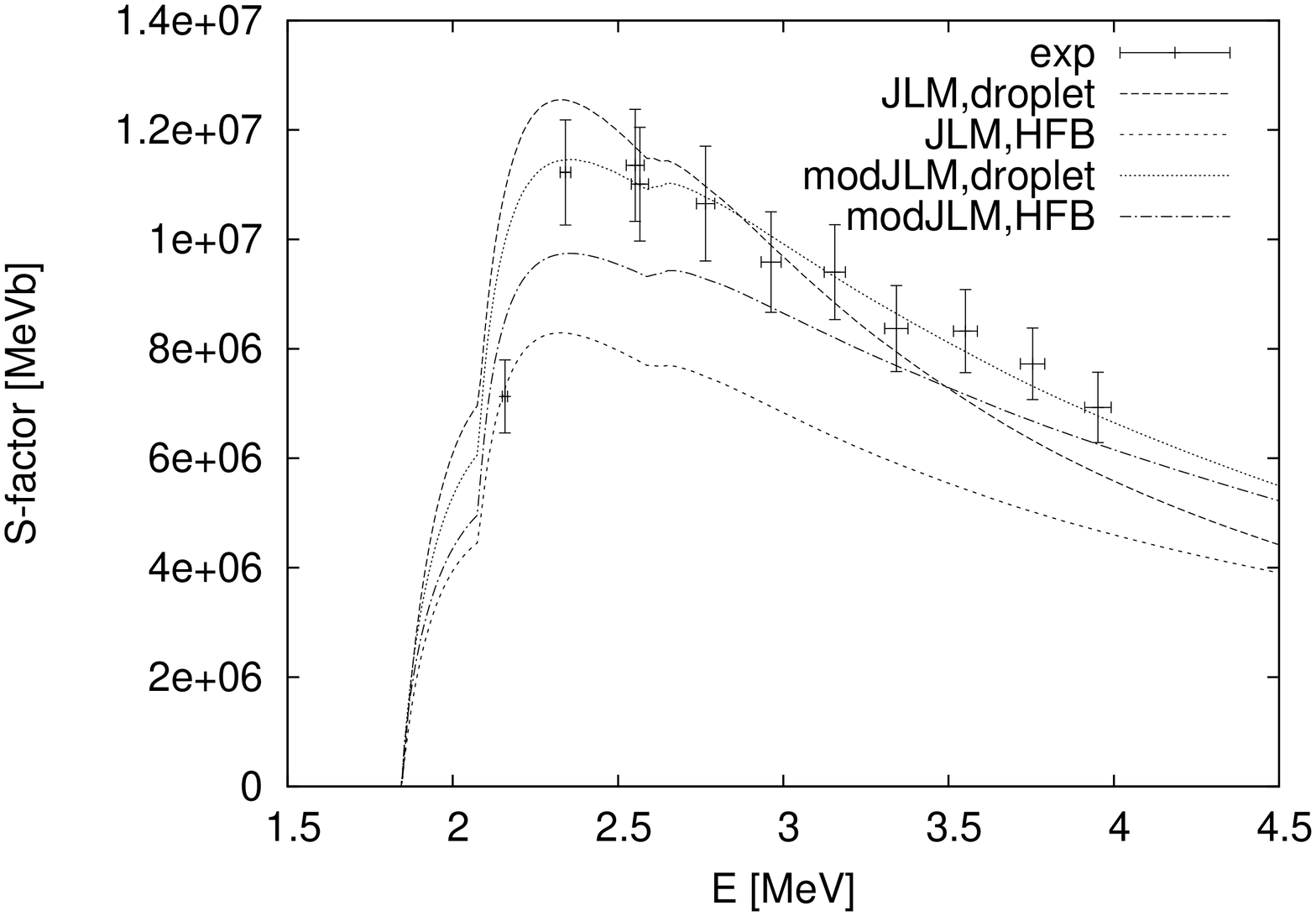}
\end{minipage}\hspace{5.5pc}%
\begin{minipage}{14pc}
\includegraphics[width=14pc,angle=-90,bb=50 100 554 770,clip=true]{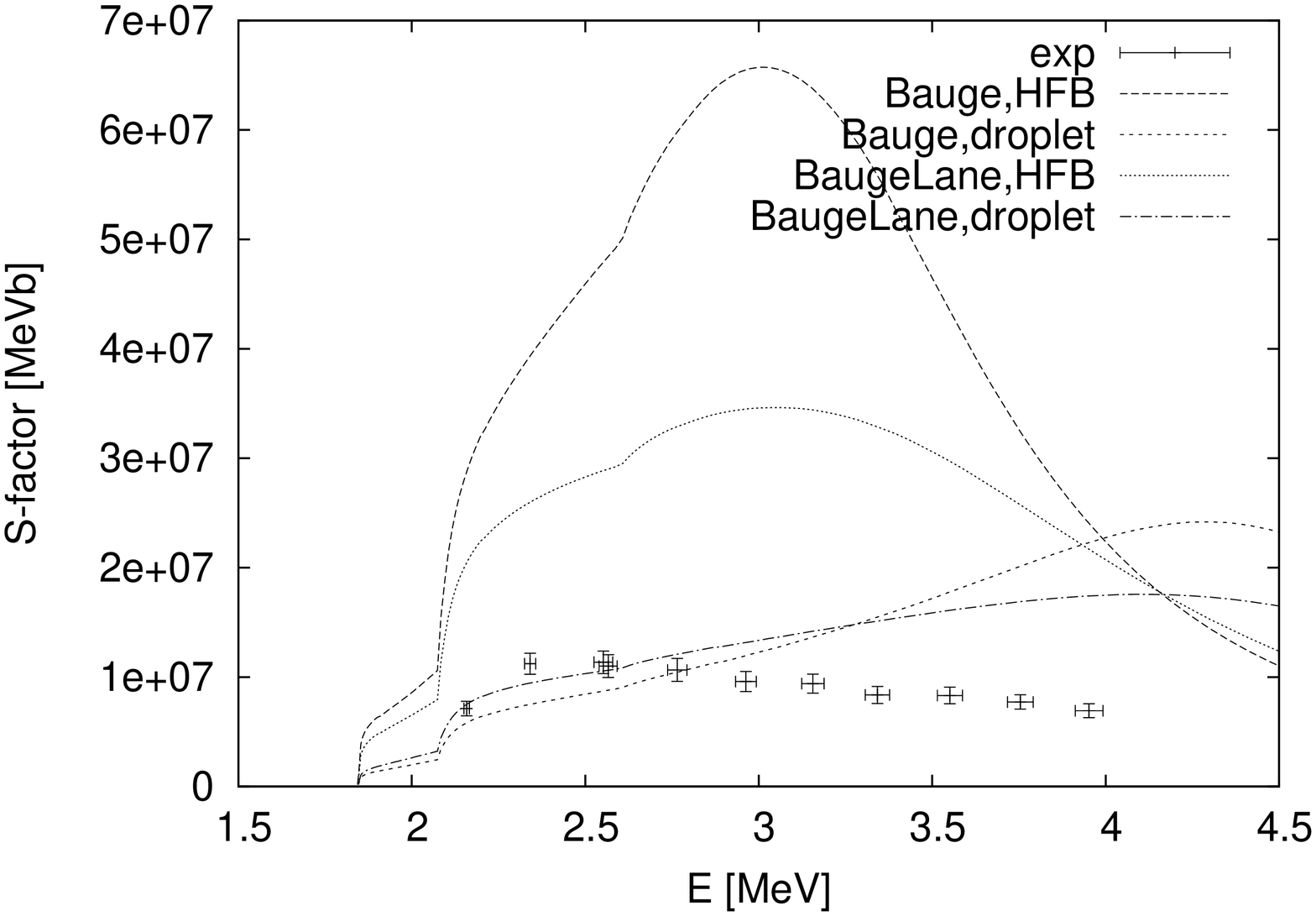}
\end{minipage} 
\caption{\label{fig:dens}Astrophysical S-factors of $^{85}$Rb(p,n)$^{85}$Sr (exp.\ data from \cite{kiss08}) compared with theory using different
optical potentials and nuclear densities. Shown are results with nuclear density from a droplet model \cite{droplet} and from a HFB model with Skyrme interaction \cite{hfb}, applied in the calculation of the "standard" potential \cite{jlm,jlmlow} and our new modified version of this potential (left panel) as well as the potential of \cite{bauge} (Bauge, right panel).}
\end{figure}

Simultaneously with the (p,$\gamma$) data, also (p,n) data is described well by our newly modified potential, as shown in \cite{prc76kiss}.
It also works well when comparing
to even more recent data (see Fig.\ \ref{fig:dens})
which was derived independently \cite{kiss08,rau09}).
 
Required input to the calculation of the optical potential is the nuclear density distribution. Fig.\ \ref{fig:dens} also shows the dependence of the results when employing a droplet model density \cite{droplet} and one from an Hartree-Fock-Bogolyubov model \cite{hfb}. For the reactions considered here, the droplet description yields better agreement to the data in both absolute scale and energy dependence of the theoretical S-factor. Therefore, all other shown results make use of this description if not explicitly mentioned otherwise. For comparison, Fig.\ \ref{fig:dens} also shows the results when employing the optical potentials of \cite{bauge98,bauge} with both densities. In the original work, HFB densities were employed \cite{bauge98,bauge}.


Summarizing, it was found that an improved reproduction of low-energy proton-induced data can be achieved by utilizing the potential of \cite{jlm,jlmlow} with a 70\% increase in the strength of the imaginary part.
This is not to say that the potential of
\cite{jlm,jlmlow} should generally be changed but rather that there is room
for a possible energy-dependent modification of the imaginary potential strength,
acting at low proton energies. This can either be achieved by adapting (refitting) the parameterization of \cite{jlm} or by externally applying an enhancement factor of the imaginary part which decreases with increasing energy.

\end{document}